    \documentstyle[nato,psfig]{crckapb}

\begin{opening}
\title{Early Galactic Evolution of Carbon, Nitrogen and Oxygen}

\author{G. Israelian$^1$}
\author{R. J. Garcia Lopez$^{1,2}$}
\author{R. Rebolo$^1$}
\institute{1. Instituto de Astrof\'\i sica de Canarias\\
           E-38200 La Laguna, Tenerife, Spain \\
           2. Departamento de Astrof\'\i sica, Universidad de La Laguna\\
	   E-38071 La Laguna, Tenerife, Spain}
	   
\end{opening}

\runningtitle{Early Galactic Evolution of Carbon, Nitrogen and Oxygen}

\begin{document}


\begin{abstract}

We present results on  carbon, nitrogen, and oxygen abundances for  a sample of
unevolved metal-poor stars with metallicities in the range $-$0.3$<$ [Fe/H]$<
-3$. Oxygen abundances derived from different indicators are compared showing
consistently that  in the range  $0.3 >$[Fe/H]$>-3.0$, the [O/Fe] ratio
increases from approximately 0 to 1. We find a good agreement between  abundances based on the forbidden line, the OH and IR triplet lines when 
gravities based on
{\it Hipparcos} parallaxes are considered for the sample stars. Gravities derived from LTE ionization balance in metal-poor stars with [Fe/H]$< -$1 
are likely too low, and could be responsible for an underestimation of
the oxygen abundances derived using the [O\,{\sc i}] line. [C/Fe] and [N/Fe] ratios  appear to be constant, independently of metallicity, in the same range. However, they  show larger scatter than oxygen at a given metallicity, which could reflect the larger variety of stellar production sites for these  other elements.

\end{abstract}

\section{Introduction}

CNO abundances in metal-poor stars  can tell us  about nucleosynthesis and 
mixing processes in stars, and about several fundamental parameters of Galactic
chemical evolution (initial mass function, star formation rate, etc.).
Unevolved stars play a key  role in this respect because the original surface
abundances of CNO in evolved giant stars can be altered by internal
nucleosynthesis and mixing between the core and outer layers.  Mixing is
expected to occur when a star becomes a red giant following the exhaustion  of
hydrogen in the core (so called ``the first dredge-up"). Important results have
been obtained from the analysis of CNO molecular bands and  atomic lines in
 metal-poor stars (see Wheeler et al. 1989 and references therein). However,
such analyses have been  either based on high resolution spectra of small
samples (less than 10) or low/intermediate resolution spectra of large samples
($\sim$ 100) of stars. We are studying CNO abundances from high resolution
spectra of more than 40 halo dwarfs. In this paper we report preliminary
results based on 24 halo dwarfs discussed by Israelian, Garc\'\i a L\'opez and
Rebolo (1998) and briefly  discuss the  current understanding of the galactic
evolution of these elements.

\section{Observations and Analysis}

The observations were carried out in different runs using the UES ($R=\lambda
/\Delta\lambda\sim 50000$) of the 4.2-m WHT at the Observatorio  del Roque de
los Muchachos (La Palma), and the UCLES ($R\sim 60000$) of the  3.9-m AAT.
The  final signal-to-noise ratio (S/N) varies for the different echelle 
orders,  being in the range 30--100  for most of the stars.

A grid of LTE, plane-parallel, constant flux, and blanketed model atmospheres 
provided  by Kurucz (1992), computed  with {\sc atlas9} without overshooting,
and interpolated for given  values of $T_{\rm eff}$, $\log g$, and  [Fe/H] was
used. Details on the analysis of the OH lines have been presented  in Israelian
et al. (1998). We have derived carbon abundances  from  the CH band at 3145
\AA\ using the list of Kurucz (1992). Nitrogen  abundances have been derived
from the NH band at 3360 \AA\ using the  atomic  and molecular  data from
Norris (1999, private communication). Synthetic spectra  were computed with the
WITA3 code by Pavlenko (1991). Effective temperatures ($T_{\rm eff}$) for our
stars were estimated using the Alonso et al. (1996) calibrations versus
$V$$-$$K$ and $b$$-$$y$ colors, which were  derived applying the infrared flux
method, and cover a wide range  of spectral types and metal content.
Metallicities were adopted from literature values obtained from high resolution
spectra.  Gravities were derived using the accurate parallaxes  measured by
{\it Hipparcos} (ESA 1997). These gravity values are larger by 0.28
dex  in average than the values adopted in our  previous analysis of OH lines
(Israelian et al. 1998). This implies a mean small reduction of 0.09 dex in the
oxygen abundances inferred in the latter paper, which does not affect the
previously  observed linear relationship between [O/Fe] and [Fe/H].

\section{Abundances of CNO in metal poor dwarfs}
\subsection{Oxygen}

Type II SNe are expected to produce significant amounts of oxygen.  Iron is
produced in both, Type II and in Type I SNe. Since the latter  come from longer
lifetime progenitors, it has been argued for a long  time that oxygen must be
overabundant in very old stars. Evidence  for high [O/Fe] ratios in many
metal-poor stars has been reported during the last decades. A so called 
``traditional'' view is based on the study of [O\,{\sc i}] lines at 6300 and
6363 \AA\ in giants (though the second line at 6363 \AA\ is not visible in very
metal-poor stars and the analysis is based {\it only on one line}) by Barbuy
(1988), Gratton \& Ortolani (1992), Sneden et al. (1991) and Kraft et al.
(1992). These authors found that [O/Fe]$=0.3-0.4$ dex at [Fe/H]$< -1$ and is
constant with decreasing metallicity. In contrast, oxygen abundances derived 
in dwarfs using the
O\,{\sc i} IR triplet at 7774 \AA\ by Abia \& Rebolo (1989), Tomkin et al.
(1992), King \& Boesgaard (1995), and Cavallo, Pilachowski, \& Rebolo (1997)
point towards increasing [O/Fe] values with decreasing [Fe/H], reaching a ratio
$\sim 1$ for stars with [Fe/H]$\sim -3$, suggesting a higher production of
oxygen during the early Galaxy.

New oxygen abundances derived from near-UV OH lines (which form in the same
layers of the atmosphere as [O\,{\sc i}]) for 24 metal-poor stars have been
presented by Israelian, Garc\'\i a L\'opez, \& Rebolo (1998). It is shown how
the [O/Fe] ratio of metal-poor stars increases from 0.6 to 1 between
[Fe/H]=$-$1.5 and $-3$ with a slope of $-0.31\pm 0.11$ (Fig 1). Contrary to the
previously accepted picture (Bessell, Sutherland, \& Ruan 1991, who used older
model atmospheres with a coarser treatment of the opacities in the UV), these
new oxygen abundances derived from low-excitation OH lines, agreed well with
those derived from high-excitation lines of the O\,{\sc i} IR triplet at 7774
\AA. The comparison with oxygen abundances derived using O\,{\sc i} data from
Tomkin et al. (1992) showed a mean difference of $0.00\pm 0.11$ dex for the
stars in common. On the other hand, Boesgaard et al. (1999) have obtained high
quality Keck spectra of many metal-poor stars in the near UV, and recently
concluded their analysis of a different set of OH lines. They find a very good
agreement with the results obtained by Israelian et al. (1998), and  basically
the same dependence of [O/Fe] versus metallicity. The mean difference in oxygen
abundance for ten stars in common is $0.00\pm 0.06$ dex when the differences in
stellar parameters are taken into account. In Fig. 1, upper panel, we plot 
these oxygen abundances based on OH lines as a function of metallicity.

\begin{figure}
\psfig{figure=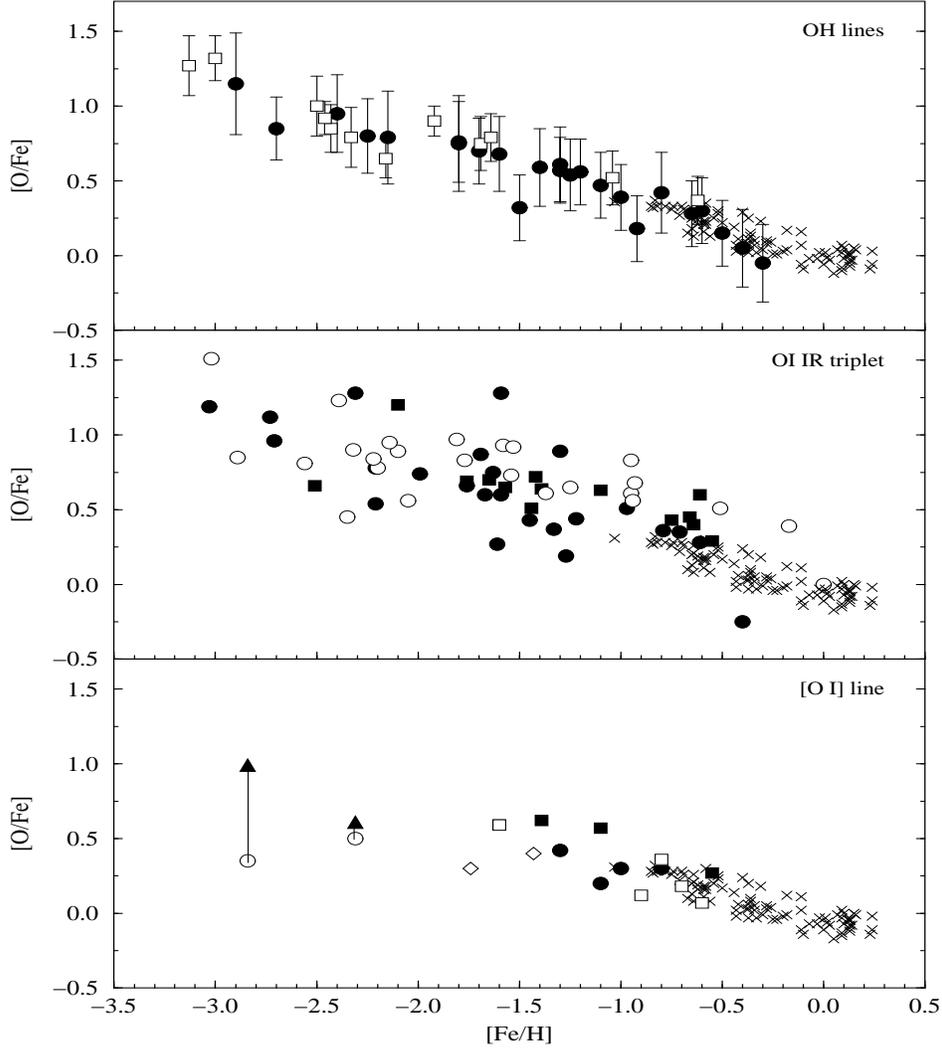,width=5in,height=6in,angle=0}
\caption{ [O/Fe] vs. [Fe/H] for unevolved stars. Abundances from OH lines
	were derived by Israelian et al. (1998; filled circles) and
	Boesgaard et al. (1999; open squares, corrected to the scale of
	stellar parameters adopted by Israelian et al.). Abundances from the 
	IR triplet were derived in NLTE by Mishenina et al. (2000; filled 
	circles), Cavallo et al. (1997; filled squares, corrected for NLTE
	effects by Mishenina et al.), and in LTE by Boesgaard et al. (1999;
	open circles). Finally, abundances from the [O\,{\sc i}] line come
	from Spiesman \& Wallerstein (1991; open diamonds), Spite \& Spite 
	(1991; open squares), Israelian et al. (1998; filled circles), 
	Mishenina et al. (2000; filled squares), and Fulbright \& Kraft (1999; 
	open circles). Filled triangles indicate the change in abundances
	associated with the change in gravity according to 
        by the {\it Hipparcos}
	parallaxes for the two stars studied by Fulbright \& Kraft. The
	abundances derived by Edvardsson et al. (1993; crosses) are shown in 
	the three plots to indicate the trend in metal-rich stars.}
\end{figure}

Balachandran \& Bell (1998) have pointed out that the continuous opacity of the
Sun is not fully accounted for in the spectral syntheses performed in the near
UV region. Although this is still a matter of  debate, it would have a minor
effect on the recent OH results since most of the stars in the samples of
Israelian et al. and Boesgaard et al. are hotter than the Sun (and very
metal-poor), and the corrections to oxygen abundances for individual stars due
to this effect would be lower than 0.15 dex. This would not affect
significantly the [O/Fe] vs. [Fe/H] trend.

Recently, Mishenina et al. (2000) performed a non-LTE analysis of the O\,{\sc
i} IR triplet to re-derive oxygen abundances for a sample of 38 metal-poor
stars  from the literature. They confirmed earlier results (Abia \& Rebolo
1989; Tomkin et al. 1992; Kiselman 1993) indicating that the mean value of the
non-LTE  correction in unevolved metal-poor stars is typically 0.1 dex and
never exceeds 0.2 dex. Mishenina et al. found the same linear trend as
Israelian et al. (1998) and Boesgaard et al. (1999) from the OH lines, and
confirmed that  oxygen abundances do not show any trend with $T_{\rm eff}$ or
$\log g$ (Boesgaard et al. 1999). Furthermore, Asplund et al. (1999) showed
that the O\,{\sc i} IR triplet is not affected by 3D effects, convection and
small-scale inhomogeneities in the stellar atmosphere. In addition, oxygen
abundances derived form this triplet are not significantly affected by
chromospheric activity
either, and we can conclude that the O\,{\sc i} IR triplet provides reliable
oxygen abundances in metal-poor dwarfs. This conclusion should also apply to
the oxygen  abundances derived from the UV OH lines given the good agreement
shown between both indicators. In Fig. 1, mid-panel, we plot oxygen abundances
based on the oxygen triplet. The larger scatter observed, as compared with the
measurements based on OH lines, can be associated with the different scales of
stellar parameters ($T_{\rm eff}$, gravities, and metallicities) adopted by the
authors of each set of stars, and to the fact that some  measurements have not
been corrected for NLTE effects.

Israelian et al. (1998) found four dwarfs (HD 22879, HD 76932,  HD
103095 and HD 134169) in their sample for which oxygen abundances had been
previously derived using [O\,{\sc i}]. They synthesized the forbidden oxygen
line for these stars adopting the same set of stellar parameters than for the
OH analysis, the $gf$ value given by Lambert (1978), and the equivalent widths
provided in the literature for the $\lambda$ 6300 \AA\ line. The estimated
abundances were in reasonable agreement with those derived from OH but still
slightly lower. The abundances found in that work using {\it Hipparcos}
gravities when analyzing both indicators are in better agreement, which
strongly  suggests that a reliable gravity scale may indeed be key to explain
the discrepancies on oxygen abundances from forbidden and permitted lines in
unevolved metal-poor stars.

In the lower panel of Fig.1 we compile oxygen measurements for unevolved stars
based on the [O\,{\sc i}] $\lambda$ 6300 \AA\ line. The presence of a linear
trend of [O/Fe] versus metallicity strongly depends on the only two
measurements available at [Fe/H]$\le -2$. These two measurements have been
recently reported by  Fulbright \& Kraft (1999) for the stars  BD $+37$ 1458
and BD $+23$ 3130, which were also considered by Israelian et al.(1998) and
Boesgaard et al. (1999; only BD $+37$ 1458 in this case).  There is an apparent
discrepancy between the results obtained from the forbidden and the OH lines.
However, we argue here that this discrepancy cannot be sustained when  a
critical analysis of the uncertainties involved in the determination  from the
forbidden line is performed. The analysis carried out by Fulbright \& Kraft  
is based on gravities derived from LTE iron ionization balance of these
subgiants where it is well known that NLTE effects are strong. In a recent
paper,  Allende Prieto et al. (1999) have shown that gravities derived using
this technique in metal-poor stars do not agree with the gravities inferred
from accurate {\it Hipparcos} parallaxes, which casts a shadow upon oxygen
abundance analyses of very metal-poor stars based on gravities derived from the
ionization balance. They find that gravities are systematically underestimated
when derived from ionization balances and that upward corrections of 0.5 dex or
even higher can be required at metallicities similar to those of our stars. We
remark here that any  underestimation of gravities will also strongly
underestimate the abundances inferred from the forbidden line. For the two
stars under discussion our {\it Hipparcos} based gravities are 0.45 and 1.05
dex (for  BD $+37$ 1458 and BD $+23$ 3130, respectively) higher than derived by
Fulbright \& Kraft, and would imply the corrections in the oxygen abundances
indicated in Fig. 1 (the details of the analysis are out of the scope of these
proceedings and  will be presented in a forthcoming paper). Our conclusion is
that the uncertainties in the gravities of these subgiants allow the abundances
inferred from the forbidden line to be consistent with those estimated from the
OH lines or the triplet. Actually, consistency with the other oxygen indicators
is achieved for the high gravities inferred from {\it  Hipparcos}, and this
could be taken as an indication that the high gravities are indeed the correct
ones.

Chemical evolution models of the early Galaxy where stellar lifetimes  are
taken into account and assuming that Type Ia SN appear at a Galactic age of 30
million years can also explain the evolution of oxygen delineated in Fig. 1.
(Chiappini et al. 1999.). The evolution of  oxygen proposed in this paper also
helps to understand the evolution  of $^6$Li versus [Fe/H] and the $^6$Li/Be
ratio at low metallicities  in the framework of standard Galactic Cosmic Ray
Nucleosynthesis (Fields \& Olive 1999). In addition, Ramaty et al. (1999) have
proposed that a delay between the effective deposition times into the ISM of 
Fe and O (only a fraction of which condensed in oxide grains) can  explain a
linear trend of [O/Fe].

\subsection{Carbon}

Until very recently, it has been commonly accepted that intermediate mass
stars  ($M_\odot < M < 10M_\odot$) are the main source of Galactic C  (Wheeler
et al. 1989). It has been shown (Laird 1985,  Carbon et al. 1987, Tomkin et al.
1992) that [C/Fe] is approximately  zero independently of [Fe/H]. Note that all
these studies were based  on the 4300 \AA\ feature of CH. Tomkin et al. (1992)
have demonstrated  that C\,{\sc i} at $\sim$ 9100 \AA\ provides an average
[C/Fe]=+0.3$\pm$0.2, whereas the CH band provides  [C/Fe]=$-$0.1$\pm$0.2.
Recently, Gustafsson et al. (1999) performed an abundance analysis of carbon in
a sample of 80 unevolved disk stars. They found that [C/Fe] increases with
decreasing [Fe/H] with a slope of $-$0.17$\pm$0.03. This result was explained
by carbon enrichment from  superwinds of metal-rich massive stars. Our
preliminary analysis for  more metal-poor stars confirms earlier results that
[C/Fe]$\sim$0 in a wide range of [Fe/H] (Fig. 2).  Studies of very low
metallicity halo stars ( [Fe/H]$< -3$) have revealed  a significant number of
stars with very high overabundance of carbon, up to 1-2 dex (Beers et al.
1992). Given these new results, it appears necessary to review our knowledge
about carbon production sites in the Galaxy.

\begin{figure}
\psfig{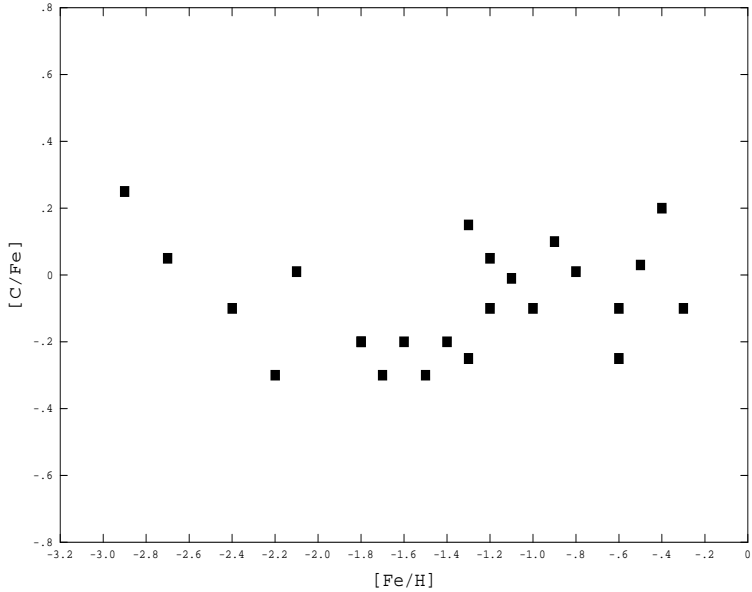}
\caption{Carbon abundances in halo dwarfs. Typical error bar is 
0.3 dex.}
\end{figure}

\subsection{Nitrogen}

The isotope $^{14}$N is synthesized from $^{12}$C and $^{16}$O through  the CNO
cycles in the H-burning layer. Observations of the NH band  at 3360 \AA\  have
allowed to delineate the Galactic evolution of N down to  [Fe/H]$\sim -$2.8.
Tomkin \& Lambert (1984) used high resolution  spectra of 8 disk and 6 halo
stars ($-0.3<$ [Fe/H]$< -2.3$) and  found [N/Fe]$\simeq -$0.25. Laird (1985)
and Carbon et al. (1987)  obtained [N/Fe]= $-0.67\pm0.14$ (intermediate
resolution  spectra of 116 stars) and [N/Fe]= $-0.11\pm0.06$ (low resolution
spectra of 76 stars), respectively. However, we should stress that  this band
is blended with Ti and Sc lines and it is preferable to  use high resolution
spectra in order to avoid any systematics due  to the overabundance of Ti
($\alpha$-element) and Sc in metal-poor stars.  Our preliminary analysis 
confirms previous results that [N/Fe] is constant in a wide range  of
[Fe/H].    It would be extremely interesting to check how [N/O] behaves in
ultra metal-poor stars with [Fe/H]$< -3$. The trend of [N/Fe] at very low
metallicities is not yet investigated but the existence of N rich halo
subdwarfs has been already demonstrated. It is very important to check the
[N/O] trend at [Fe/H]$<-$3 as this may  help to understand the possible sites
for a production of nitrogen  as  primary element. Recently, Maeder \& Meynet
(2000) have shown that the average rotation of massive stars can possibly be
faster at lower metallicities. They have also found a primary N production  in
rapidly rotating stars in the mass range 10-15 $M_{\odot}$. To our knowledge 
these are the only models showing a primary nitrogen production in normal
stars (see also Maeder, A., this conference).

\begin{figure}
\psfig{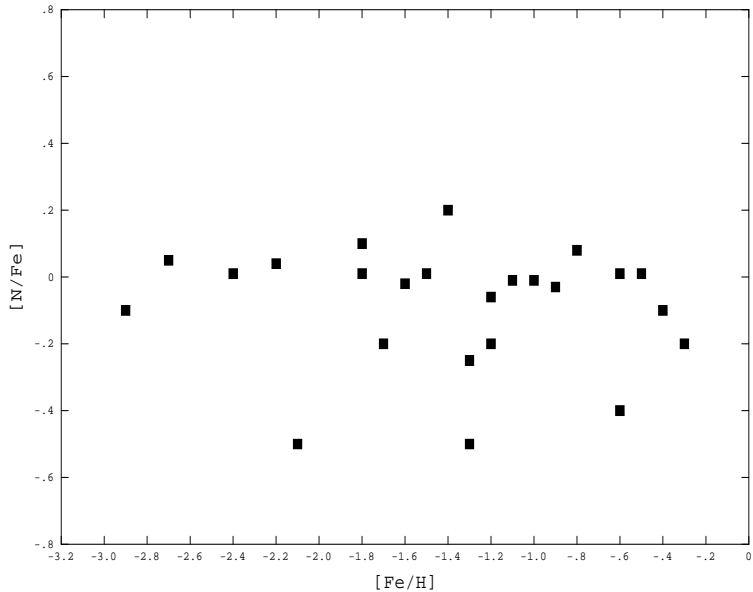}
\caption{Nitrogen abundances in halo dwarfs. Typical error bar is
0.2 dex.}
\end{figure}

\section{Concluding remarks}

We have derived carbon, nitrogen and oxygen abundances for a sample of
metal-poor unevolved stars. [C/Fe] and [N/Fe] ratios  appear to be constant,
independently of metallicity, in the range  $0.3 >$[Fe/H]$>-3.0$, while the
[O/Fe] ratio increases from approximately 0 to 1, with consistent oxygen
abundances derived from different indicators. Carbon and nitrogen abundances
show larger scatter than oxygen at a given metallicity, which could reflect the
larger variety of stellar production sites for these elements.  Work is in
progress to derive CNO abundances for a larger sample of unevolved metal-poor
stars using other features in the optical and IR spectral regions, which should
allow a fully reliable estimate of the Galactic evolution of these elements.
Consistent abundances from different spectral features of C, N and O will be
used in order to delineate [N/O], [C/O] and [(C+N+O)/Fe] behaviour as a
function of [O/H] and [Fe/H].

\end{document}